\documentclass[final,5p,times,onecolumn]{elsarticle}
\usepackage{pifont}
\usepackage{amsmath}
\usepackage{amsfonts}
\usepackage{url}
\usepackage{graphics}
\usepackage{graphicx}
\usepackage{mathrsfs}
\usepackage{amssymb}
\usepackage{amsthm}
\usepackage{float}
\biboptions{sort&compress}

\newtheorem{lemma}{Lemma}
\newtheorem{remark}{Remark}
\newtheorem{assumption}{Assumption}
\newtheorem{theorem}{Theorem}
\newtheorem{definition}{Definition}

\journal{Neurocomputing}

\begin{document}

\begin{frontmatter}

\title{Adaptive Neural Control for a Class of Stochastic Nonlinear Systems with Unknown Parameters, Unknown Nonlinear Functions and Stochastic Disturbances}
\author[label0,label1]{Chao-Yang Chen\corref{cor1}}
\cortext[cor1]{Corresponding author.}
\ead{cychen@hnust.edu.cn}
\author[label1]{Wei-Hua Gui}
\author[label3]{Zhi-Hong Guan}
\author[label2]{Ru-Liang Wang}
\author[label0]{Shao-Wu Zhou}

\address[label0]{School of Information and Electrical Engineering, Hunan University of Science and Technology, Xiangtan, 411201, P.R. China}
\address[label1]{School of Information Science and Engineering, Central South University, Changsha, 410012, P.R.China}
\address[label3]{College of Automation, Huazhong University of Science and Technology, Wuhan, 430074, P. R. China}
\address[label2]{Computer and Information Engineering College, Guangxi Teachers Education University, Nanning, 530001, P.R. China}

\begin{abstract}
~~~In this paper, adaptive neural control (ANC) is investigated for a class of strict-feedback nonlinear stochastic systems with unknown parameters, unknown nonlinear functions and stochastic disturbances. The new controller of adaptive neural network with state feedback is presented by using a universal approximation of radial basis function neural network and backstepping. An adaptive neural network state-feedback controller is designed by constructing a suitable Lyapunov function. Adaptive bounding design technique is used to deal with the unknown nonlinear functions and unknown parameters. It is shown that, the global asymptotically stable in probability can be achieved for the closed-loop system. The simulation results are presented to demonstrate the effectiveness of the proposed control strategy in the presence of unknown parameters, unknown nonlinear functions and stochastic disturbances.
\end{abstract}


\end{frontmatter}


\section{Introduction}
In recent years, the study of the robust control system design for nonlinear systems has attracted extensive attention. Lots of significant developments have been obtained\cite{1,2,3,4,4a,4b,4c,4d,4d1,4d2,4f,4g}, and interesting results of adaptive nonlinear control have been ever-increasing. Adaptive backstepping as a powerful method has a large number of applied research about synthesizing controllers for lower-triangular nonlinear systems. Backstepping design technique is one of the methods to design nonlinear system by structuring intermediate laws and Lyapunov functions step by step. It has obtained a large number of successful applications in nonlinear control. Such as \cite{6,7,8}.
\par
Adaptive control is an important branch of robust control. Noticeably, due to neural network control and fuzzy logic control have a good approximation ability over a compact domain, they are very suitable for handle highly uncertain and nonlinear system, and they have become an important part of adaptive control and a lot of research has been obtained, such as \cite{5,9,10,11}. In the early development stage of neural network control schemes, the control schemes that derive parameter adaptive law in off-line environments was usually used \cite{12}, which can perform well in some simple cases, but the stability, robustness and performance of nonlinear systems have few systematic analytical methods. In order to avoid the above problems, the reference \cite{13} has obtain some adaptive neural control schemes based on Lyapunov's stability theory.
\par
Stochastic nonlinear modelling has come to play an important role in many areas of industry, science and technology. After the success of systematic control design for deterministic nonlinear systems, how to extend this technique to the case of stochastic nonlinear systems has been an open research area \cite{14}-\cite{23}. Therefore, it is a challenging and meaningful issue fo the stability analysis and control design of nonlinear stochastic systems, and have attracted more and more scholars' attention in recent years. The main technical obstacle in the Lyapunov function design for stochastic nonlinear systems is that the gradient and the higher order Hessian term are involved $It\hat{o}$ stochastic differentiation. In \cite{14,15}, strict-feedback stochastic system is studied for the first time using  a backstepping design. \cite{16} tried to extend the results in \cite{14}, a class stochastic nonlinear systems with time-delay is investigated. By using the quadratic Lyapunov function, \cite{17,18} studied the stabilization problem of stochastic nonlinear systems. In \cite{17}, a class of stochastic nonholonomic system has investigated, and adaptive stabilization by state-feedback is resolved. For linear LTI SISO plants, \cite{19} proposed a modified adaptive backstepping control. In \cite{16}, a class of stochastic nonlinear systems were investigated via output feedback, linearly bounded unmeasurable states are involved in nonlinear function. By an adaptive neural control (ANC) scheme, \cite{30a} studied a class of non-affine pure feedback stochastic nonlinear system. It is shown that all the signals involved are semi-globally uniformly ultimately bounded under the action of the developed controller. A simplified adaptive backstepping neural control (ABNC) strategies are proposed for a class of uncertain strict-feedback nonlinear systems in \cite{30b}. In \cite{36}, a class of stochastic non-strict-feedback stochastic nonlinear system is studied. However, the system functions must are the monotonously increasing and it should be bounded, which is a stricter assumption because it is the difficulty to confirm a bounded and monotonously increasing unknown function. In \cite{37}, this constraints that bounded and monotonously increasing for nonlinear functions are relaxed, but it is unavailable when the stochastic system with unknown parameters and unknown disturbances. And then, a series of studies on stochastic systems are carried out \cite{21,22,23}.
\par
At the same time, the ANNC method has been many successful applications for some unknown nonlinear systems, such as adaptive output-feedback control \cite{26,24,25}, pure-feedback \cite{27,29,28,30} and so on. In ANNC, the neural network is often used to online approximate unknown nonlinearity owning to their inherent approximation capabilities. Motivated by the above observations, this work focuses on the feedback nonlinear stochastic systems with unknown parameters, unknown nonlinear functions and stochastic disturbances, using the adaptive neural control method. By using backstepping method, an adaptive neural controller is designed. The proposed adaptive neural controller guarantees that all the signals in the closed-loop are bounded. Partial results of this paper have been presented in \cite{31}.
\par
In this paper, the main contributions in this paper are as follows:
~(i)~The new controller of adaptive neural network with state feedback is presented by using a universal approximation of radial basis function neural network and backstepping for a class of stochastic nonlinear systems. Compared with the results in \cite{37a,37b}, the gain function for virtual control signal $x_{i+1}$ is constant 1. In this work, it is extended to a function of the partial state variables. Therefore, the systems in \cite{37a,37b} are special situations of the system considered for this work.
~(ii)~Corresponding model is more universal than references \cite{30a,30b,37}. The unknown disturbances and unknown nonlinear functions are considered. Adaptive bounding design technique is used to deal with the unknown parameters and unknown nonlinear functions. The upper bound of the disturbances is not necessary to know in the design of the adaptive controller. When the derivatives of the state are available for feedback, the designed adaptive controller can be guarantee that all the signals of the closed-loop system are bounded;
~(iii)~A universal type adaptive feedback controller is designed, which globally regulates all the states of the uncertain system while keeping boundedness of all the states to guarantee global asymptotically stable in probability. The main advantage of the proposed control scheme is that the all closed-loop signals are guaranteed to be globally bounded. However, many existing adaptive backstepping neural control approaches (e.g., see \cite{5,9,11,13,19,24,25,28,30,30a,33,34}) can only guarantee semi-globally bounded for all the closed-loop signals.
\section{Problem formulation and preliminaries}
\subsection{Plant dynamics}

In this section, we recall the basic background knowledge concerning the stability properties of stochastic systems.
\par
Consider the following stochastic nonlinear system
\begin{eqnarray}\label{eqq1}
       \text{d}x_i&=&(g_i(\bar{x}_i)x_{i+1}+{\theta_{i}^*}^T\psi_{i}^*(\bar{x}_i)+f_i(\bar{x}_i)+\Delta_i(x,t))\text{d}t\nonumber\\
       &&+\phi_i^T(\bar{x}_i)\text{d}\omega~~~1\leq~i\leq~n-1,\nonumber\\
       \text{d}{x}_n&=&(g_n(\bar{x}_n)u+{\theta_{n}^*}^T\psi_{n}^*(\bar{x}_n)+f_n(\bar{x}_n)+\Delta_n(x,t))\text{d}t\nonumber\\
       &&+\phi_n^T(\bar{x}_n)\text{d}\omega,\nonumber\\
       y(t)&=&x_1(t),\hfill
\end{eqnarray}
where $x=[x_1,x_2,\cdots,x_n]^T\in{R^n}$ is the state vector with initial value $x(0)$, $\bar{x}_i=[x_1,x_2,\ldots,x_i]$; $u\in{R}$ is the control input and $y\in{R}$ is the system output, respectively; $f_i(\cdot)\in{R}$ and $\psi_i^*(\cdot)\in{R^{q_i}},\psi_i(0)=0$ are unknown smooth functions with $f_i(0)=0, \psi_i^*(0)=0$; $g_i(\cdot)\in{R}, g_i(\cdot)\neq{0}$ and $\phi_i(\cdot)\in{R^r}$ are known smooth functions, ${\theta_i^*}\in{R}^{q_i}$ are unknown bounded parameters,
$i=1,\ldots,n$; $\Delta_i(x,t)$ is unknown disturbances. $\omega$ is an independent $r$-dimensional standard Wiener process and $\Delta_i(x,t)$ satisfy the following assumption.
\begin{assumption}\label{ass1}\cite{33}
For unknown disturbances $\Delta_i(x,t)$ and $\forall{(t,x)}$ $\in{R^+}\times{R^n}$, we have
\begin{equation}
|\Delta_i(x,t)|\leq{p_i^*\Phi_i^*(\bar{x}_i)}\label{eqq2}
\end{equation}
where $p_i^*\geq0$ and $\Phi_i(\bar{x_i})$ are some unknown parameter values and known smooth functions, respectively. And, $\Phi_i^*(0)=0$, we define $p_i^*$ is the smallest nonnegative constant such that (\ref{eqq2}) is satisfied.
\end{assumption}
To simplify the notation, we let $\theta=[{\theta_{1}^*}^T,\cdots,{\theta_{n}^*}^T]^T\in{R^q}$,
where $q:=\sum_iq_i$. Therefore, we can rewrite the system $(\ref{eqq1})$ as in following form:
\begin{align}
       \text{d}x_i~=&~(g_i(\bar{x}_i)x_{i+1}+\theta^T\Psi_i(\bar{x}_i)+f_i(\bar{x}_i)+\Delta_i(x,t))\text{d}t\nonumber\\
       &+\phi_i^T(\bar{x}_i)\text{d}\omega~~~1\leq~i\leq~n-1,\nonumber\\
       \text{d}{x}_n~=&~(g_n(\bar{x}_n)u+\theta^T\Psi_n(\bar{x}_n)+f_n(\bar{x}_n)+\Delta_n(x,t))\text{d}t\nonumber\\
       &+\phi_n^T(\bar{x}_n)\text{d}\omega,\nonumber\\
       y(t)~=&~x_1(t),\hfill\label{eqq3}
\end{align}
where each $\Psi_i: R^i\mapsto{R^q}$ is given by
\begin{align*}
\Psi_i=&[\Psi_{i1}, \Psi_{i2},\cdots, \Psi_{in}]\\
=&[0_1^T,\cdots,0_{i-1}^T,\psi_i^*,0_{i+1}^T,\cdots,0_{n}^T]^T
\end{align*}
with $0_i:=[0,\cdots,0]^T\in{R^q_i}$.

\begin{assumption}\label{ass2}
For unknown functions $\Psi_{ij}(\bar{x}_i)$ and $\forall{x}$ $\in{R^n}$, we have
\begin{equation}
|\Psi_{ij}(\bar{x}_i)|\leq{b_{ij}^*\varphi_{ij}^*(\bar{x}_i)}\label{eq3}
\end{equation}
where $\varphi_{i}^*=[\varphi_{i,1}^*,\varphi_{i,2}^*,\cdots,\varphi_{i,nq}^*]^T$, $\varphi_{i}^*(0)=0$. And $b_{ij}^*\geq0,$ is unknown parameter values and $\varphi^*_{ij}(\bar{x}_i)$ is known smooth functions, respectively. $b_{ij}^*$ is defined as the smallest nonnegative constant such that (\ref{eqq2}) is satisfied.
\end{assumption}

\begin{remark}
Consider the system (\ref{eqq3}) can not satisfy either the linear parametrization conditions, such as following system
\begin{align*}
       \text{d}x_1~=&~((x_1^2+1)x_1x_{2}+\theta_1x_1^2+\theta_3\sin{(t\theta_4x_2)})\text{d}t+x_1\text{d}\omega,\\
       \text{d}{x}_2~=&~(e^{x_2}u+\theta_1x_2^2+\theta_2e^{x_1}+(\theta_4+\theta_3\sin{x_1})x_2^2)
       +e^{-x_2})\text{d}t,\\
       y(t)~=&~x_1(t).\hfill
\end{align*}
It can be put in the form (\ref{eqq2}) by letting
\begin{eqnarray}
\begin{array}{lll}
  g_1=x_1^2+1,g_2=e^{-x_2},& \theta=[\theta_1~~\theta_2]^T,\\[5pt]
  f_1=0,f_2=e^{-x_2}, & \phi_1=x_1,\phi_2=0,\\[5pt]
  \Psi_1=[x_1^2~~0]^T, & \Psi_2=[x_2^2~~e^{x_1}]^T,\\[5pt]
  \Delta_1=\theta_3\sin(t\theta_4x_2), & \Delta_2=(\theta_4+\theta_3\sin{x_1})x_2^2.
\end{array}\nonumber
\end{eqnarray}
The bounds on $|\Delta_1|\leq{p_1^*},|\Delta_1|\leq{p_1^*x_2^2}$, where
\[p_1:=|\theta_3|, p_2:=|\theta_3|+|\theta_4|.\]
\end{remark}
\begin{definition}\cite{38,Khasminskii}
If for any $\varepsilon>0$,
\[\lim_{x(0)\rightarrow0}\limits{P}\Bigg\{\sup_{t\geq{0}}\limits\parallel{x(t)}||\geq\varepsilon\Bigg\}=0,\]
and for any initial condition $x(0)$, we have $P\Big\{\lim_{t\rightarrow\infty}\limits{x(t)}=0\Big\}=1$, then
we claim that the solution $\{x(t)=0\} $ of system (\ref{eqq3}) is asymptotically stable in the large.
\end{definition}
\begin{definition}\cite{Khasminskii}
If
\[\lim_{\varepsilon\rightarrow\infty}\limits\sup_{t\geq{0}}\limits{P}\Big\{\|x(t)\|\geq\varepsilon\Big\}=0.\]
The solution process $\{x(t),t\geq0\}$ of system (\ref{eqq3}) is said to be bounded in probability.
\end{definition}
Denote $L$ with
\begin{equation}\label{eqq4}
LV(x)=\frac{\partial{V}}{\partial{x}}f+\frac{1}{2}Tr\Big\{{g^T}\frac{\partial^2{V}}{\partial{x}^2}g\Big\},
\end{equation}
the infinitesimal generator of the solution of stochastic system (\ref{eqq3}) for any $V(x)\in{C^2(R^n;R)}$.
\begin{lemma}\label{lem1}\cite{34}
For system $(\ref{eqq3})$, if there exists a positive definite radial unbounded
Lyapunov functions $V_1(x)\in{C^2(R^n;R)}$ and $V_2(\theta)\in{C^2(R^m;R)}$,
and $V(x,\theta)=V_1(x)+V_2(\theta)$ satisfying
\begin{equation}
LV(x)\leq-\lambda{V}(x)+K,\label{eqq5}
\end{equation}
where constants $\lambda>0,~K\geq0$, then, the stochastic system $(\ref{eqq3})$ is globe bounded stable in probability.
\end{lemma}
\begin{lemma}\label{lem2}\cite{35}
For any $\epsilon>0$ and $u\in{R}$, the following inequality holds
\begin{equation*}
0\leq|u|-u\tanh(u/\epsilon)\leq\delta\epsilon,
\end{equation*}
where $\delta=0.2785.$
\end{lemma}
\subsection{RBFNN Approximation}
In this work, the RBFNN is is used to approximate a nonlinear continuous function. The continuous function $Q(Z):R^q\rightarrow R$ is given by
\begin{align}\label{5}
Q_{nn}(Z,W)=W^TS(Z),
\end{align}
where $Z\in \Omega\subset R^q$ and,
$$W=[w_1,w_2,\ldots,w_l]^T\in R^l,$$ are the input vector and weight vector, respectively. Denote the number of nodes in the neural network is $l>1$; and
$$S(Z)=[s_1(Z),\ldots,s_l(Z)]^T,$$
with $s_i(Z)$ being chosen as the following form Gaussian functions
\begin{equation}s_i(Z)=exp\left[\frac{-(Z-\mu_i)^T(Z-\mu_i)}{\eta_i^2}\right],\,(i=1,2,\ldots,l.)\nonumber\end{equation}
where $\mu_i=[\mu_1,\mu_2,\ldots,\mu_q]^T$ and $\eta$ represent the center of the
receptive field and the width of the Gaussian function, respectively. Is well known that neural network $(\ref{5})$ can approximate any continuous function defined on a compact set $\Omega_Z\subset R^q$, furthermore, the neural network can is given by
\begin{equation}
Q(Z)=Q_{nn}(Z,W^*)+\varepsilon(Z)~~~\forall\in\Omega_Z,\label{eqq6}
\end{equation}
where $W^*$ is the ideal neural network weights and
$|\varepsilon(Z)|\leq\varepsilon^*$ is the neural network approximation
error. $W^*$ is ideal constant weights, and for all $Z\in\Omega_Z$, there exist constant $\varepsilon^*>0$ such that $|\varepsilon|\leq\varepsilon^*$. Moreover, $W^*$ is bounded by $\|W^*\|\leq w_m$ on the compact set $\Omega_Z$, where $w_m$ is a positive constant.
\par
$W^*$ is usually unknown and need to be estimated
in function approximation. $W^*$ is defined as follows:
\begin{equation}
W^*=arg\min_{(W)}\limits\left[\sup_{Z\in\Omega_Z}|Q_{nn}(Z,W)-Q(Z)|\right],\nonumber
\end{equation}
which is unknown and needs to be estimated in control design. Let
$\hat{W}$ be the estimate of $W^*$, and let $\tilde{W}=\hat{W}-W^*$
be the weight estimation error.
\section{Adaptive control design and stability analysis} 
In this section, adaptive control design is given by backstepping design,
and there contain $n$ steps in backstepping design procedure. At each step,
an intermediate control function $\alpha_i(t)$ shall be developed
using an appropriate Lyapunov function $V_i(t)$. The design of both
the control laws and the adaptive laws are based on the following
change of coordinates:
\begin{equation}
z_1=x_1,\,z_i=x_i-\alpha_{i-1},\,i=2,\ldots,n,\label{eqq7}
\end{equation}
where $\alpha_i(t)$ is an intermediate control. And, the system controller $u(t)$ will is designed in the last step to stabilize the entire closed-loop system.
\par
Step 1:~~~
The Lyapunov function candidate is consider as
\begin{equation}\label{eqq8}
V_1=\frac{z_1^4}{4}+\frac{\tilde{\vartheta}_1^T\Gamma^{-1}_{\vartheta_1}\tilde{\vartheta}_1}{2}+\frac{\tilde{p}_1^T\Gamma_{p_1}\tilde{p}_1}{2}
+\frac{\tilde{W}_1^T\Gamma^{-1}_{w_1}\tilde{W}_1}{2}+\frac{\tilde{\varepsilon}_1^2}{2\gamma_{\varepsilon_1}},
\end{equation}
where $\Gamma^{-1}_{\vartheta_1}$ and $\Gamma_{w_1}$ are symmetric positive definite matrices, $\Gamma_{p_1}\leq{0}, \Gamma_{\varepsilon_1}\leq{0}.$
Let $\hat{\vartheta}_i$ is the $i$th times estimate of $\vartheta$, for clarify in every step, denote $\vartheta_i=\vartheta$ for the $i$th step.
In order to facilitate of analysis, we denote $g_i(\bar{x}_i(t))=g_i$, others may be deduced by analogy,
Then, we have
\begin{eqnarray}
\text{d}z_1=(g_1(z_2+\alpha_1)+\theta_1^T\Psi_1+f_1+\Delta_1(x,t))\text{d}t+\phi_1^T\text{d}\omega,\label{eqq9}
\end{eqnarray}
Let $\Lambda_1=\Delta_1, p_1=p_1^*, \Phi_1=\Phi_1^*, \varphi_1=\varphi_1^*$.
By using (\ref{eqq7}) and (\ref{eqq9}), we have
\begin{eqnarray}
\text{d}z_1=(g_1(z_2+\alpha_1)+\theta_1^T\Psi_1+f_1+\Lambda_1(x,t))\text{d}t+\phi_1^T\text{d}\omega.\label{eqq10}
\end{eqnarray}
Note (\ref{eqq4})(\ref{eqq8}) and (\ref{eqq10}), we can obtain
\begin{align}
LV_1~=&~z_1^3(g_1(z_2+\alpha_1)+\theta_1^T\Psi_1+f_1+\Lambda_1(x,t))\nonumber\\
&-\tilde{\theta}^T\Gamma^{-1}_{\theta_1}\dot{\hat{\theta}}-\tilde{p}_1^T\Gamma^{-1}_{p_1}\dot{\hat{p}}_1
-\gamma^{-1}_{\varepsilon_1}\tilde{\varepsilon}_1^T\dot{\hat{\varepsilon}}_1
-\tilde{W}_1^T\Gamma^{-1}_{w_1}\dot{\hat{W}}_1\nonumber\\
&+3z_1^2\phi_1^T\phi_1/2.\label{eqq11}
\end{align}
Let $Q_1(Z_1)=f_1$.
Applying Yong's inequalities, we have
\begin{align}
3z_1^2\phi_1^T\phi_1/2~\leq&~\frac{3\epsilon_{11}}{4}+\frac{3z_1^4}{4\epsilon_{11}}\parallel\phi_1\parallel^4,\label{eqq12}\\
g_1z_1^3z_2~\leq&~\frac{3}{4}g_1^{4/3}z_1^4+\frac{1}{4}z_{2}^4,\label{eqq13}
\end{align}
and applying Assumption (\ref{ass2}), we have
\begin{align}\label{eq15}
\theta_1^T\Psi_1
&\leq\sum_{k=1}^{q}|\theta_{1k}|\varphi_{1k}^*\nonumber\\
&\leq\sum_{k=1}^{q}|\theta_{1k}|b_{1k}\varphi_{1k}^*\nonumber\\
&\leq\vartheta_{1}^T\varphi_{1}^*\nonumber\\
&=\vartheta_{1}^T\varphi_{1}
\end{align}
where
$$\vartheta_{1}=[|\theta_{11}|b_{11},\cdots,|\theta_{1q}|b_{1q}]^T, ~ \varphi_{1}^*=[\varphi_{11}^*,\cdots,\varphi_{1q}^*]^T.$$
From (\ref{eqq11})-(\ref{eq15}) and applying Assumption \ref{ass1}, we have
\begin{align}
LV_1~\leq&~\frac{3}{4}g_1^{4/3}z_1^4+\frac{1}{4}z_2^4+z_1^3g_1\alpha_1+|z_1|^3\vartheta_1^T\varphi_1+z_1^3{W_1^*}^TS(Z_1)\nonumber\\
&+|z_1|^3\varepsilon_1^*+|z_1|^3p_1\Phi_1(x_1)-\tilde{\theta}^T\Gamma^{-1}_{\theta_1}\dot{\hat{\theta}}
-\tilde{p}_1^T\Gamma^{-1}_{p_1}\dot{\hat{p}}_1\nonumber\\
&-\gamma^{-1}_{\varepsilon_1}\tilde{\varepsilon}_1^T\dot{\hat{\varepsilon}}_1
-\tilde{W}_1^T\Gamma^{-1}_{w_1}\dot{\hat{W}}_1+\frac{3}{4}\epsilon_{11}\nonumber\\
&+\frac{3z_1^4}{4\epsilon_{11}}\parallel\phi_1\parallel^4.
\label{eqq14}
\end{align}
Let intermediate law as
\begin{align*}
\alpha_1~=&~g_1^{-1}(-c_1z_1-\frac{3}{4}g_1^{4/3}z_1-\beta_{10}-\beta_{11}-\beta_{12}\\
&-\hat{W}_1^TS(Z_1)-\frac{3z_1}{4\epsilon_{11}}\parallel\phi_1\parallel^4),\\
\beta_{10} ~=&~\hat{\varepsilon}_1\varpi_{10}, \varpi_{10}:=\tanh[\frac{z_1^3}{\varepsilon_{10}}],\\
\beta_{11} ~=&~ \hat{p}_1^T\varpi_{11}, \varpi_{11}:=\Phi_1(\bar{x}_1)\tanh[\frac{z_1^3\Phi_1(\bar{x}_1)}{\varepsilon_{11}}];\\
\beta_{12} ~=&~ \hat{\vartheta}_1^T\varpi_{12}, \varpi_{12}:=\varphi_1(\bar{x}_1)\tanh[\frac{z_1^3\varphi_1(\bar{x}_1)}{\varepsilon_{12}}];
\end{align*}
Then, we have
\begin{align}
LV_1~\leq&~-c_1z_1^4+\frac{1}{4}z_2^2-\tilde{\vartheta}_1^T(\Gamma_{\vartheta_1}^{-1}\dot{\hat{\vartheta}}_1-z_1^3\varphi_1)\nonumber\\
&-\tilde{\varepsilon}_1(\gamma_{\varepsilon_1}^{-1}\dot{\hat{\varepsilon}}_1+z_1^3\varpi_{10})
-\tilde{p}_1^T(\Gamma_{p_1}^{-1}\dot{\hat{p}}_1+z_1^3\varpi_{11})\nonumber\\
&-\tilde{W}_1^T(\Gamma_{w_1}^{-1}\dot{\hat{W}}_1+z_1^3S(Z_1))+|z_1|^3\varepsilon_1^*-z_1^3\varepsilon_1^*\varpi_{10}\nonumber\\
&+|z_1|^3p_1^T\Phi_1(x_1)-z_1^3p_1^T\varpi_{11}+|z_1|^3\vartheta_1^T\varphi_1(x_1)-z_1^3\vartheta_1^T\varpi_{12}\nonumber\\
&+\frac{3}{4}\epsilon_{11}.\label{eqq17}
\end{align}
The adaptive laws are chosen as
\begin{align}
\begin{split}
\dot{\hat{\vartheta}}_1~=&~-\Gamma_{\vartheta_1}(z_1^3\varphi_1-\sigma_{\vartheta_1}\hat{\vartheta}_1),~(\sigma_{\vartheta_1}>0)\\
\dot{\hat{\varepsilon}}_1~=&~-\gamma_{\varepsilon_1}(z_1^3\varpi_{10}-\sigma_{\varepsilon_1}\hat{\varepsilon}_1),~(\sigma_{\varepsilon_1}>0)\\
\dot{\hat{p}}_1~=&~-\Gamma_{p_1}(z_1^3\varpi_{11}-\sigma_{p_1}\hat{p}_1),~(\sigma_{p_1}>0)\\
\dot{\hat{W}}_1~=&~-\Gamma_{w_1}(z_1^3S(Z_1)-\sigma_{w_1}\hat{W}_1),~(\sigma_{w_1}>0)
\end{split}\label{eqq18}
\end{align}
Note (\ref{eqq17}) and Lemma (\ref{lem2}), we have
\begin{align}
LV_1~\leq&~-c_1z_1^4+\frac{1}{4}z_2^2+\sigma_{\vartheta_1}\tilde{\vartheta}_1^T\hat{\vartheta}_1
+\sigma_{\varepsilon_1}\tilde{\varepsilon}_1\hat{\varepsilon}_1+\sigma_{p_1}\tilde{p_1}^T\hat{p}_1\nonumber\\
&+\sigma_{W_1}\tilde{W}_1^T\hat{W}_1+0.2785(\varepsilon_{10}+\varepsilon_{11}+\varepsilon_{12})+\frac{3}{4}\epsilon_{11}.\label{eqq19}
\end{align}
Applying following inequalities
\begin{align}
\begin{split}
\sigma_{\vartheta_1}\tilde{\vartheta}_1^T\hat{\vartheta}_1~\leq&~-\frac{1}{2}\sigma_{\vartheta_1}\parallel\tilde{\vartheta}_1\parallel^2+\frac{1}{2}\sigma_{\vartheta_1}\parallel\vartheta_1\parallel^2,\\
\sigma_{\varepsilon_1}\tilde{\varepsilon}_1\hat{\varepsilon}_1~\leq&~-\frac{1}{2}\sigma_{\varepsilon_1}\parallel\tilde{\varepsilon}_1\parallel^2+\frac{1}{2}\sigma_{\varepsilon\varepsilon_1}\|\varepsilon_1\|^2,\\
\sigma_{p_1}\tilde{p}_1\hat{p}_1~\leq&~-\frac{1}{2}\sigma_{p_1}\parallel\tilde{p}_1\parallel^2+\frac{1}{2}\sigma_{p_1}\parallel{p_1}\parallel^2,\\
\sigma_{w_1}\tilde{W}_1^T\hat{W}_1~\leq&~-\frac{1}{2}\sigma_{w_1}\parallel\tilde{W}_1\parallel^2+\frac{1}{2}\sigma_{w_1}\parallel{W_1^*}\parallel^2,
\end{split}\label{eqq20}
\end{align}
we can obtain
\begin{align}
LV_1~\leq&~-c_1z_1^4+\frac{1}{4}z_2^4-\frac{1}{2}\sigma_{\vartheta_1}\parallel\tilde{\vartheta}_1\parallel^2
-\frac{1}{2}\sigma_{\varepsilon_1}\parallel\tilde{\varepsilon}_1\parallel^2\nonumber\\
&-\frac{1}{2}\sigma_{p_1}\parallel\tilde{p}_1\parallel^2-\frac{1}{2}\sigma_{w_1}\parallel\tilde{W}_1\parallel^2
+\frac{1}{2}\sigma_{p_1}\parallel{p_1}\parallel^2\nonumber\\
&+\frac{1}{2}\sigma_{\vartheta_1}\parallel\vartheta_1\parallel^2+\frac{1}{2}\sigma_{w_1}\parallel{W_1^*}\parallel^2+\frac{3}{4}\epsilon_{11}\nonumber\\
&+0.2785(\varepsilon_{10}+\varepsilon_{11}+\varepsilon_{12})\nonumber\\
\leq&~-\lambda_1{V_1}+K_1+\frac{1}{4}z_{2}^4,\label{eqq21}
\end{align}
where
\begin{align*}
\lambda_1=&\min\left\{4c_1,\gamma_{\varepsilon_1\sigma_{\varepsilon_1}},
\frac{\sigma_{\vartheta_1}}{\max(\Gamma_{\vartheta_1}^{-1})},
\frac{\sigma_{p_1}}{\max(\Gamma_{p_1}^{-1})},\frac{\sigma_{w_1}}{\max(\Gamma_{w_1}^{-1})}\right\},\\
K_1=&\frac{1}{2}\sigma_{p_1}\parallel{p_1}\parallel^2
+\frac{1}{2}\sigma_{\vartheta_1}\parallel\vartheta_1\parallel^2
+\frac{1}{2}\sigma_{w_1}\parallel{W_1^*}\parallel^2\\
&+\frac{3}{4}\epsilon_{11}+0.2785(\varepsilon_{10}+\varepsilon_{11}+\varepsilon_{12}).
\end{align*}
\par
Step $i$:~~~$(2\leq{i}<n)$. Similarly,
we can obtain
\begin{align}
\text{d}z_i~=&~\left[g_i(z_{i+1}+\alpha_i)+\theta_i^T\Psi_i+f_i+\Delta_i(x,t)\right]\text{d}t+\phi_i\text{d}\omega\nonumber\\
&-\left[\sum_{j=1}^{i-1}\frac{\partial{\alpha}_{i-1}}{\partial{x}_j}(g_jx_{j+1}+\theta_j^T\Psi_j+f_j+\Delta_j)\right.\nonumber\\
&+\sum_{j=1}^{i-1}\frac{\partial{\alpha}_{i-1}}{\partial{\hat{\vartheta}_j}}\dot{\hat{\vartheta}}_j
+\sum_{j=1}^{i-1}\frac{\partial{\alpha}_{i-1}}{\partial{\hat{W}_j}}\dot{\hat{W}}_j
+\sum_{j=1}^{i-1}\frac{\partial{\alpha}_{i-1}}{\partial{\hat{\varepsilon}_j}}\dot{\hat{\varepsilon}}_j\nonumber\\
&+\sum_{j=1}^{i-1}\frac{\partial{\alpha}_{i-1}}{\partial{\hat{p_j}}}\dot{\hat{p}}_j
+\left.\frac{1}{2}\sum_{k,j=1}^{i-1}\frac{\partial^2{\alpha}_{i-1}}{\partial{{x}_k}\partial{{x}_j}}g_kg_j^T\right]\text{d}t\nonumber\\
&-\sum_{j=1}^{i-1}\frac{\partial{\alpha}_{i-1}}{\partial{x_j}}\phi_j\text{d}\omega.\label{eqq22}
\end{align}
Applying Yong¡¯s inequalities, we have
\begin{align}
\frac{3z_i^2}{2}(\phi_i-\sum_{j=1}^{i-1}&\frac{\partial{\alpha}_{i-1}}{\partial{x_j}}\phi_j)
(\phi_i-\sum_{j=1}^{i-1}\frac{\partial{\alpha}_{i-1}}{\partial{x_j}}\phi_j)^T\nonumber\\
\leq&~\frac{3\epsilon_{i1}}{4}+\frac{3z_i^4}{4\epsilon_{i1}}\parallel{\phi_i-\sum_{j=1}^{i-1}\frac{\partial{\alpha}_{i-1}}{\partial{x_j}}\phi_j}\parallel^4,\label{eqq23}\\
g_iz_i^3z_{i+1}~\leq&~\frac{3}{4}g_1^{4/3}z_i^4+\frac{1}{4}z_{i+1}^4.\label{eqq24}
\end{align}
Let
\begin{align}
Q_i(Z_i)&=f_i-\sum_{j=1}^{i-1}\frac{\partial{\alpha}_{i-1}}{\partial{x_j}}f_j(\bar{x}_j),~~~(2\leq{i}\leq{n})\label{eqq25}\\
|\Lambda_i|&=|\Delta_i-\sum_{j=1}^{i-1}\frac{\partial\alpha_{i-1}}{\partial{x_j}}\Delta_j|,~~~(2\leq{i}\leq{n})\label{eqq27}\\
|\Upsilon_i|&=|\theta_i^T\Psi_i-\sum_{j=1}^{i-1}\frac{\partial\alpha_{i-1}}{\partial{x_j}}\theta_i^T\Psi_j|,~~~(2\leq{i}\leq{n})\label{eqq27a}\\
V_i=&V_{i-1}+\frac{z_i^4}{4}+\frac{\tilde{\vartheta}^T\Gamma^{-1}_{\vartheta_i}\tilde{\vartheta}}{2}+\frac{\tilde{p}_i^T\Gamma_{p_i}\tilde{p}_i}{2}+\frac{\tilde{W}_i^T\Gamma^{-1}_{w_i}\tilde{W}_i}{2}+\frac{\tilde{\varepsilon}_i^2}{2\gamma_{\varepsilon_i}},\label{eqq26}
\end{align}
Define operator $|\theta_i|_1=[|\theta_{i1}|,|\theta_{i2}|,\cdots,|\theta_{iq}|]^T$, then using assumption (\ref{ass1}) and (\ref{ass2}), we have
\begin{align}
|\Lambda_i|\leq{p_i}^T\bar{\Phi}_i,~
|\Upsilon_i|\leq\vartheta_i^T\bar{\varphi}_i,\label{eqq28}
\end{align}
where
\begin{align}
p_i~=&~[p_1^*,p_2^*,\cdots,p_i^*]^T,\nonumber\\
\bar{\Phi}_i ~=&~\Big[|\frac{\partial\alpha_{i-1}}{\partial{x_1}}|\Phi_1^*,
|\frac{\partial\alpha_{i-1}}{\partial{x_2}}|\Phi_2^*,\cdots,|\frac{\partial\alpha_{i-1}}{\partial{x_{i-1}}}|\Phi_{i-1}^*,\Phi_i^*\Big]^T,\nonumber\\
\vartheta_i~=&~\Big[(|\theta_1|_1{\bullet}b_1^*)^T,(|\theta_2|_1{\bullet}b_2^*)^T,\cdots,(|\theta_i|_1{\bullet}b_i^*)^T\Big]^T,\nonumber\\
\bar{\varphi}_i ~=&~\Big[|\frac{\partial\alpha_{i-1}}{\partial{x_1}}|{\varphi_1^*}^T,
|\frac{\partial\alpha_{i-1}}{\partial{x_2}}|{\varphi_2^*}^T,\cdots,|\frac{\partial\alpha_{i-1}}{\partial{x_{i-1}}}|{\varphi_{i-1}^*}^T,{\varphi_i^*}^T\Big]^T.\nonumber
\end{align}
Let
\begin{align}
\Phi_i~=&~[\frac{\partial\alpha_{i-1}}{\partial{x_1}}\Phi_1^*,
\frac{\partial\alpha_{i-1}}{\partial{x_2}}\Phi_2^*,\cdots,\frac{\partial\alpha_{i-1}}{\partial{x_{i-1}}}\Phi_{i-1}^*,\Phi_i^*]^T,\nonumber\\
\varphi_i ~=&~\Big[\frac{\partial\alpha_{i-1}}{\partial{x_1}}{\varphi_1^*}^T,
\frac{\partial\alpha_{i-1}}{\partial{x_2}}{\varphi_2^*}^T,\cdots,\frac{\partial\alpha_{i-1}}{\partial{x_{i-1}}}{\varphi_{i-1}^*}^T,{\varphi_i^*}^T\Big]^T,\nonumber
\end{align}
with $2\leq{i}\leq{n}$.\\
By using (\ref{eqq22})--(\ref{eqq28}), we have
\begin{align}
LV_i~\leq&~LV_{i-1}+\frac{3}{4}g_i^{4/3}+\frac{1}{4}z_{i+1}^4+z_i^3g_i\alpha_i\nonumber\\
&+|z_i|^3\vartheta_i^T\bar{\varphi}_i + z_i^3{W_i^*}^TS(Z_i)\nonumber\\
&+|z_i|^3\varepsilon_i^*+|z|_i^3p_i^T\bar{\Phi}_i-z_i^3\sum_{j=1}^{i-1}(\frac{\partial{\alpha}_{i-1}}{\partial{x}_j}x_{j+1}
\nonumber\\
&+\frac{\partial\alpha_{i-1}}{\partial{\hat{\vartheta}_j}}\dot{\hat{\vartheta}}_j+\frac{\partial\alpha_{i-1}}{\partial{\hat{W}_j}}\dot{\hat{W}}_j
+\frac{\partial\alpha_{i-1}}{\partial{\hat{\varepsilon}_j}}\dot{\hat{\varepsilon}}_j
+\frac{\partial\alpha_{i-1}}{\partial{\hat{p}_j}}\dot{\hat{p}}_j)\nonumber\\
&-\frac{z_i^3}{2}\sum_{k,j=1}^{i-1}\frac{\partial^2\alpha_{i-1}}{\partial{x_k}\partial{x}_j}g_kg_j^T
+\frac{3\epsilon_{i1}}{4}\nonumber\\
&+\frac{3z_i^4}{4\epsilon_{i1}}\parallel{\phi_i-\sum_{j=1}^{i-1}\frac{\partial{\alpha}_{i-1}}{\partial{x_j}}\phi_j}\parallel^4
-\tilde{\vartheta}_i^T\Gamma_{\vartheta_i}^{-1}\dot{\hat{\vartheta}}_i\nonumber\\
&-\tilde{W}_i^T\Gamma_{W_i}^{-1}\dot{\hat{W}}_i
-\tilde{p}_i^T\Gamma_{p_i}^{-1}\dot{\hat{p}}_i-\gamma_{\varepsilon_i}^{-1}\tilde{\varepsilon}_i\dot{\hat{\varepsilon}}_i.\label{eqq29}
\end{align}
We select intermediate law $\alpha_i,~2\leq{i}\leq{n-1}$ as
\begin{align}
\alpha_i~=&~g_i^{-1}\Bigg[-c_iz_i-\frac{1}{4}z_i-\frac{3}{4}g_1^{4/3}z_i-\beta_{i0}-\beta_{i1}-\beta_{i2}\nonumber\\
&-\hat{W}_i^TS(Z_i)
+\frac{1}{2}\sum_{k,j=1}^{i-1}\frac{\partial^2{\alpha}_{i-1}}{\partial{{x}_k}\partial{{x}_j}}g_kg_j^T\nonumber\\
&+\sum_{j=1}^{i-1}(\frac{\partial{\alpha}_{i-1}}{\partial{x}_j}x_{j+1}+\frac{\partial{\alpha}_{i-1}}{\partial{\hat{\vartheta}}_j}\dot{\hat{\vartheta}}_j
+\frac{\partial{\alpha}_{i-1}}{\partial{\hat{W}}_j}\dot{\hat{W}}_j+\frac{\partial\alpha_{i-1}}{\partial{\hat{\varepsilon}_j}}\dot{\hat{\varepsilon}}_j\nonumber\\
&+\frac{\partial\alpha_{i-1}}{\partial{\hat{p}_j}}\dot{\hat{p}}_j)
-\left.\frac{3z_i}{4\epsilon_{i1}}\parallel\phi_i-\sum_{j=1}^{i-1}\frac{\partial{\alpha}_{i-1}}{\partial{x_j}}\phi_j\parallel^4\right].\label{eqq30}
\end{align}
Let
\begin{align}
\begin{split}
  \beta_{i0} & =  \hat{\varepsilon}_i\varpi_{i0}, ~~~\,\varpi_{i0}:=\tanh[\frac{z_i^3}{\varepsilon_{i0}}],\\
  \beta_{i1} & =  \hat{p}_i^T\varpi_{i1}, ~~\varpi_{i1}:=\Phi_i\odot\tanh[\frac{z_i^3\Phi_i}{\varepsilon_{i1}}],\\
  \beta_{i2} & =  \hat{\vartheta}_i^T\varpi_{i2}, ~~\varpi_{i2}:=\varphi_i\odot\tanh[\frac{z_i^3\varphi_i}{\varepsilon_{i1}}],
\end{split}\label{eqq31}
\end{align}
where
\begin{align*}
&\Phi_i\odot\tanh[\frac{z_i^3\Phi_i}{\varepsilon_{i1}}]\\
=&\left[\frac{\partial\alpha_{i-1}}{\partial{x_1}}\Phi_1^*\tanh[\frac{z_i^3\frac{\partial\alpha_{i-1}}{\partial{x_1}}\Phi_1^*}{\varepsilon_{i1}}],
\cdots, \frac{\partial\alpha_{i-1}}{\partial{x_{i-1}}}\Phi_{i-1}^*\tanh[\frac{z_i^3\frac{\partial\alpha_{i-1}}{\partial{x_{i-1}}}\Phi_{i-1}^*}{\varepsilon_{i1}}],\right.\\
&\left.\Phi_i^*\tanh[\frac{z_i^3\Phi_i^*}{\varepsilon_{i1}}]\right]^T,\\
&\varphi_i\odot\tanh[\frac{z_i^3\varphi_i}{\varepsilon_{i2}}]\\
=&\left[\frac{\partial\alpha_{i-1}}{\partial{x_1}}\varphi_1^*\tanh[\frac{z_i^3\frac{\partial\alpha_{i-1}}{\partial{x_1}}\varphi_1^*}{\varepsilon_{i2}}],
\cdots,\frac{\partial\alpha_{i-1}}{\partial{x_{i-1}}}\varphi_{i-1}^*\tanh[\frac{z_i^3\frac{\partial\alpha_{i-1}}{\partial{x_{i-1}}}\varphi_{i-1}^*}{\varepsilon_{i2}}],\right.\\
&\left.\varphi_i^*\tanh[\frac{z_i^3\varphi_i^*}{\varepsilon_{i2}}]\right]^T.
\end{align*}
and using Lemma (\ref{lem2}) we can obtain
\begin{align}
LV_i\leq&LV_{i-1}-\frac{1}{4}z_i^4-c_iz_i^4
-\tilde{W}_i^{T}(\Gamma_{w_i}^{-1}\dot{\hat{W}}_i+z_i^3S(Z_i))\nonumber\\
&-\tilde{\varepsilon}_i(\gamma_{\varepsilon_i}^{-1}\dot{\hat{\varepsilon}}_i+z_i^3\varpi_{i0})
-\tilde{p}_i^{T}(\Gamma_{p_i}^{-1}\dot{\hat{p}}_i+z_i^3\varpi_{i1})\nonumber\\
&-\tilde{\vartheta}_i^{T}\left[\Gamma_{\vartheta_i}^{-1}\dot{\hat{\vartheta}}_i
+z_i^3\varpi_{i2}\right]\nonumber\\
&+0.2785(\varepsilon_{i0}+\varepsilon_{i1}+\varepsilon_{i2})+\frac{3\epsilon_{i1}}{4}+\frac{1}{4}z_{i+1}^4.\label{eqq32}
\end{align}
The adaptive laws are chosen as
\begin{align}
\begin{split}
\dot{\hat{\varepsilon}}_i~=&~-\gamma_{\varepsilon_i}(z_i^3\varpi_{i0}-\sigma_{\varepsilon_i}\hat{\varepsilon}_i), ~ (\sigma_{\varepsilon_i}>0),\\
\dot{\hat{p}}_i~=&~-\Gamma_{p_i}(z_i^3\varpi_{i1}-\sigma_{p_i}\hat{p}_i), ~ (\sigma_{p_i}>0),\\
\dot{\hat{\vartheta}}_i~=&~-\Gamma_{\vartheta_i}(z_i^3(\varpi_{i2}-\sigma_{\vartheta_i}\hat{\vartheta}_i), ~ (\sigma_{\vartheta_i}>0),\\
\dot{\hat{W}}_i~=&~-\Gamma_{w_i}(z_i^3S(Z_i)-\sigma_{w_i}\hat{W}_i), ~(\sigma_{w_i}>0).
\end{split}\label{eqq33}
\end{align}
Note (\ref{eqq20}), (\ref{eqq32}), (\ref{eqq33}) can be written as
\begin{align}
LV_i~\leq&~LV_{i-1}-\frac{1}{4}z_i^4-c_iz_i^4+\sigma_{\vartheta_i}\tilde{\vartheta}_i^T\hat{\vartheta}_i
+\sigma_{\varepsilon_i}\tilde{\varepsilon}_i\hat{\varepsilon}_i\nonumber\\
&+\sigma_{p_i}\tilde{p_i}^T\hat{p}_i+\sigma_{W_i}\tilde{W}_i^T\hat{W}_i+0.2785(\varepsilon_{i0}+\varepsilon_{i1}+\varepsilon_{i2})\nonumber\\
&+\frac{3}{4}\epsilon_{i1}+\frac{1}{4}z_{i+1}^4\label{eqq34}\\
\leq&~LV_{i-1}-c_iz_i^4+\frac{1}{4}z_{i+1}^4-\frac{1}{2}\sigma_{\vartheta_i}\parallel\tilde{\vartheta}_i\parallel^2
-\frac{1}{2}\sigma_{\varepsilon_i}\parallel\tilde{\varepsilon}_1\parallel^2\nonumber\\
&-\frac{1}{2}\sigma_{p_i}\parallel\tilde{p}_i\parallel^2-\frac{1}{2}\sigma_{w_i}\parallel\tilde{W}_i\parallel^2
+\frac{1}{2}\sigma_{p_i}\parallel{p_i}\parallel^2\nonumber\\
&+\frac{1}{2}\sigma_{\vartheta_i}\parallel\vartheta_i\parallel^2+\frac{1}{2}\sigma_{w_i}\parallel{W_i^*}\parallel^2
+\frac{3}{4}\epsilon_{i1}\nonumber\\
\leq&~(LV_{i-1}-\frac{1}{4}z_i^4)-\lambda_1{V_i}+K_i+\frac{1}{4}z_{i+1}^4\nonumber\\
\leq&~-\sum_{j=1}^{i}\lambda_j{V_j}+\sum_{j=1}^{i}K_j+\frac{1}{4}z_{i+1}^4,
\label{eqq35}
\end{align}
where
\begin{align*}
\lambda_i=&\min\left\{4c_i,\gamma_{\varepsilon_i\sigma_{\varepsilon_i}},
\frac{\sigma_{\vartheta_i}}{\max(\Gamma_{\vartheta_i}^{-1})},
\frac{\sigma_{p_i}}{\max(\Gamma_{p_i}^{-1})},\frac{\sigma_{w_i}}{\max(\Gamma_{w_i}^{-1})}\right\},\\
K_i=&\frac{1}{2}\sigma_{p_i}\parallel{p_i}\parallel^2
+\frac{1}{2}\sigma_{\vartheta_i}\parallel\vartheta_i\parallel^2
+\frac{1}{2}\sigma_{w_i}\parallel{W_i^*}\parallel^2\\
&+\frac{3}{4}\epsilon_{i1}+0.2785(\varepsilon_{i0}+\varepsilon_{i1}+\varepsilon_{i2}).
\end{align*}
\par
Step $n$:~We select the control
\begin{align}
u=&g_n^{-1}\Bigg[-c_nz_n-\frac{1}{4}z_n-\beta_{n0}-\beta_{n1}-\beta_{n2}\nonumber\\
&-\hat{W}_n^TS(Z_n)
+\frac{1}{2}\sum_{k,j=1}^{n-1}\frac{\partial^2{\alpha}_{n-1}}{\partial{{x}_k}\partial{{x}_j}}\phi_k\phi_j^T\nonumber\\
&+\sum_{j=1}^{n-1}(\frac{\partial{\alpha}_{n-1}}{\partial{x}_j}x_{j+1}+\frac{\partial{\alpha}_{n-1}}{\partial{\hat{\vartheta}}_j}\dot{\hat{\vartheta}}_j
+\frac{\partial{\alpha}_{n-1}}{\partial{\hat{W}}_j}\dot{\hat{W}}_j+\frac{\partial\alpha_{n-1}}{\partial{\hat{\varepsilon}_j}}\dot{\hat{\varepsilon}}_j\nonumber
\end{align}
\begin{align}
&+\frac{\partial\alpha_{n-1}}{\partial{\hat{p}_j}}\dot{\hat{p}}_j)
-\left.\frac{3z_n}{4\epsilon_{n1}}\parallel\phi_n-\sum_{j=1}^{n-1}\frac{\partial{\alpha}_{n-1}}{\partial{x_j}}\phi_j\parallel^4\right].\label{eqq36}
\end{align}
A similar procedure is employed, by using $(\ref{eqq23}-\ref{eqq28},\ref{eqq31},\ref{eqq33})$, we have
\begin{equation*}\label{eqq37}
LV_n\leq-\sum_{i=1}^{n}\lambda_i{V_i}+\sum_{i=1}^{n}K_i,
\end{equation*}
where
\begin{align*}
\lambda_i=&\min\left\{4c_i,\gamma_{\varepsilon_i\sigma_{\varepsilon_i}},
\frac{\sigma_{\vartheta_i}}{\max(\Gamma_{\vartheta_i}^{-1})},
\frac{\sigma_{p_i}}{\max(\Gamma_{p_i}^{-1})},\frac{\sigma_{w_i}}{\max(\Gamma_{w_i}^{-1})}\right\},\\
K_i=&\frac{1}{2}\sigma_{p_i}\parallel{p_i}\parallel^2
+\frac{1}{2}\sigma_{\vartheta_i}\parallel\vartheta_i\parallel^2
+\frac{1}{2}\sigma_{w_i}\parallel{W_i^*}\parallel^2\\
&+\frac{3}{4}\epsilon_{i1}+0.2785(\varepsilon_{i0}+\varepsilon_{i1}+\varepsilon_{i2}),
\end{align*}
with $1\leq{i}\leq{n}$.
\par
At the present stage, ANC design has been completed based on backstepping technique. The main result can be summarized by the following theorem.
\begin{theorem}
Consider stochastic nonlinear systems (\ref{eqq1}), assumptions (\ref{ass1}) and (\ref{ass2}). We can design state-feedback adaptive controller (\ref{eqq36}) and adaptation laws (\ref{eqq18}) and (\ref{eqq33}) by using backstepping approach. For bounded initial conditions, the following prosperities hold.
\begin{enumerate}
  \item[(I)] The closed-loop system's the equilibrium at the origin is asymptotically stable in probability.
  \item[(II)] The closed-loop system has an almost surely unique solution on $[0,\infty)$ for each $x(0), \hat{\vartheta}(0), \hat{W}(0), \hat{\varepsilon}(0), \hat{p}(0)$, system state $x(t)$ and the parameter estimates $\hat{\vartheta}(t), \hat{W}(t), \hat{\varepsilon}(t), \hat{p}(t)$ satisfy
\begin{align*}
P\Big\{\lim_{t\rightarrow\infty}&x(t)=0\Big\}=1;\\
P\Big\{ \lim_{t\rightarrow\infty}&\hat{\vartheta}(t),
\lim_{t\rightarrow\infty}\hat{W}(t),
\lim_{t\rightarrow\infty}\hat{\varepsilon}(t)~and~
\lim_{t\rightarrow\infty}\hat{p}(t)\\
&\hspace{2cm}~exist~and~are~finite\Big\}=1.\nonumber
\end{align*}
\end{enumerate}
\end{theorem}

\section{Simulation studies}

The following example is given to show the effectiveness of the proposed adaptive neural control algorithms for stochastic nonlinear system (\ref{eqq38}). we consider the following second-order system.
\begin{align}\label{eqq38}
\begin{split}
\text{d}x_1~=&~(x_2+f_1(x_1)+\Delta_1(x,t))\text{d}t+x_1\cos{x_1}\text{d}\omega,\\
\text{d}{x}_2~=&~((1+0.5\sin{x_1})u+\theta_2^*\psi_2^*+f_2(x))\text{d}t\\
&~+\sin{x_2}\text{d}\omega,\\
y~=~&x_1,
\end{split}
\end{align}
where
\begin{align*}
\begin{array}{lllll}
   g_1=1,~~\theta_1^*\varphi_1^*=0, & f_1=x_1\sin{x_1}, & \Delta_1=0.5x_1\sin{x_2t}, \\[5pt]
   \phi_1=x_1\cos{x_1}, & g_2=(1+0.5\sin{x_1}), & \theta_2^*\psi_2^*=0.02x_2, \\[5pt]
   f_2=x_2\cos{x_2}, & \Delta_2=0, & \phi_2=\sin{x_2}.
\end{array}
\end{align*}
The adaptive laws can be designed in the following:
\begin{align*}
\begin{array}{l}
\dot{\hat{\varepsilon}}_1~=~\gamma_{\varepsilon_1}(z_1^3\varpi_{10}-\sigma_{\varepsilon_1}\hat{\varepsilon}_1), \\[5pt]
\dot{\hat{p}}_1~=~\Gamma_{p_1}(z_1^3\varpi_{11}-\sigma_{p_1}\hat{p}_1),\\[5pt]
\dot{\hat{W}}_1~=~\Gamma_{w_1}(z_1^3S(Z_1)-\sigma_{w_1}\hat{W}_1),\\[5pt]
\dot{\hat{\vartheta}}_1~=~\Gamma_{\vartheta_1}(z_1^3\varphi_1-\sigma_{\vartheta_1}\hat{\vartheta}_1),\\[5pt]
\dot{\hat{\varepsilon}}_2~=~\gamma_{\varepsilon_2}(z_2^3\varpi_{20}-\sigma_{\varepsilon_2}\hat{\varepsilon}_2),\\[5pt]
\dot{\hat{p}}_2~=~\Gamma_{p_2}(z_2^3\varpi_{21}-\sigma_{p_2}\hat{p}_2),\\[5pt]
\dot{\hat{W}}_2~=~\Gamma_{w_2}(z_2^3S(Z_2)-\sigma_{w_2}\hat{W}_2),\\[5pt]
\dot{\hat{\vartheta}}_2~=~\Gamma_{\vartheta_2}(z_2^3\varpi_{22}-\sigma_{\vartheta_2}\hat{\vartheta}_2).
\end{array}
\end{align*}
Finally, the system control law can be designed as
\begin{align*}
u~=&~g_2^{-1}\left[-c_2z_2-\frac{1}{4}z_n-\beta_{20}-\beta_{21}\right.-\hat{\vartheta}_2^T(\varphi_2
-\frac{\partial{\alpha}_{1}}{\partial{x}_1}\varphi_1)\\
&-\hat{W}_2^TS(Z_2)+\frac{1}{2}\frac{\partial^2{\alpha}_{1}}{\partial{{x}_1^2}}\phi_1^2
+(\frac{\partial{\alpha}_{1}}{\partial{x}_1}x_{2}+\frac{\partial{\alpha}_{1}}{\partial{\hat{\vartheta}}_1}\dot{\hat{\vartheta}}_1
+\frac{\partial{\alpha}_{1}}{\partial{\hat{W}}_1}\dot{\hat{W}}_1\\
&+\frac{\partial\alpha_{1}}{\partial{\hat{\varepsilon}_1}}\dot{\hat{\varepsilon}}_1+\frac{\partial\alpha_{1}}{\partial{\hat{p}_1}}\dot{\hat{p}}_1)
-\left.\frac{3z_2}{4\epsilon_{21}}(\phi_2-\frac{\partial{\alpha}_{1}}{\partial{x_1}}\phi_1)^4\right],
\end{align*}
where
\begin{align*}
\omega_{20}~=&~\tanh\Big[z_2^3/\varepsilon_{20}\Big],\\
\omega_{21}~=&~\left(
\begin{array}{c}
  \frac{\partial\alpha_1}{\partial{x_1}}\phi_1^*
  \tanh\Big[z_2^3\frac{\partial\alpha_1}{\partial{x_1}}\phi_1^*/\varepsilon_{21}\Big]\\[5pt]
  0
\end{array}\right).
\end{align*}
The initial conditions and design parameters are given as following.
\[
\begin{array}{ll}
\hat{\theta}_1(0)~=~[0~0.1]^T,&\hat{\theta}_2(0)~=~[0~0.8]^T,\\[5pt]
\hat{b}_1=[0~1]^T,&\hat{b}_2=[0~0.1]^T,\\[5pt]
\hat{\varepsilon}_1(0)~=~0.1\text{e-3},&\hat{\varepsilon}_2(0)~=~0,~\hat{p}_1(0)~=~0.1,\\[5pt]
\hat{p}_2(0)~=~[0~0.15]^T,&\hat{W}_1(0)~=~\hat{W}_1(0)~=~0,\\[5pt]
\Gamma_{\vartheta_1}~=~diag(0.3~0.3),&\Gamma_{\vartheta_2}~=~diag(0.25~0.25),\\[5pt]
\gamma_{\varepsilon_1}~=~0.3,~\gamma_{\varepsilon_2}~=~0.4,&\Gamma_{p_1}~=~0.3,~\Gamma_{p_2}~=~[0.4~0],\\[5pt]
\sigma_{{\vartheta}_1}~=~0.3,~\sigma_{{\vartheta}_2}~=~0.25,&\sigma_{{\varepsilon}_1}~=~0.3,~\sigma_{{\varepsilon}_1}~=~0.4,\\[5pt]
\sigma_{{p}_1}~=~0.3,~\sigma_{{p}_2}~=~0.4,&\sigma_{{w}_1}~=~1.5,~\sigma_{{w}_2}~=~0.3,\\[5pt]
\varepsilon_{10}~=~\varepsilon_{20}~=~\varepsilon_{11}~=~\varepsilon_{21}~=~0.3,& c_1~=~c_2~=~0.3.
\end{array}
\]
Specifically, neural network $\hat{W}^T_1S(Z_1)$ contains 27 nodes (i.e. $l_1=27$) with
centers $\mu_l(l=1,\ldots,l_1)$ evenly spaced in $[-1.5,1.5]$ and widths
$\eta_l=0.8(l=1,\ldots,l_1)$. Neural networks $\hat{W}_2^TS(Z_2)$ contains 64 nodes(i.e. $l_2=64$) with centers $\mu_l(l=1,\ldots,l_2)$ evenly spaced in $[-1.5,1.5]\times[-1.5,1.5]\times[-1.5,1.5]\times[-1.5,1.5],$ and widths
$\eta_l=1.5(l=1,\ldots,l_1)$.
\par
\begin{figure}[H]
\centering
  \includegraphics[width=8cm]{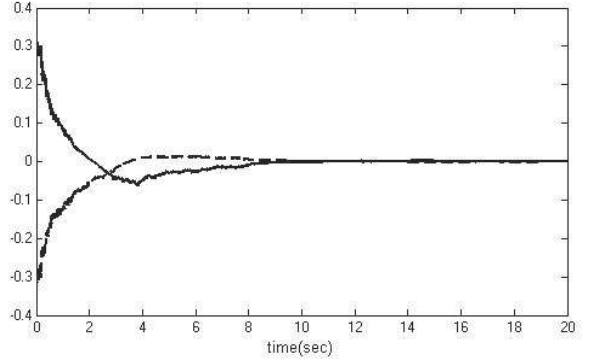}
  \caption{System states $x_1$:``solid line." $x_2$:``dash line".}\label{Fig.1}
\end{figure}
\begin{figure}[H]
\centering
  \includegraphics[width=8cm]{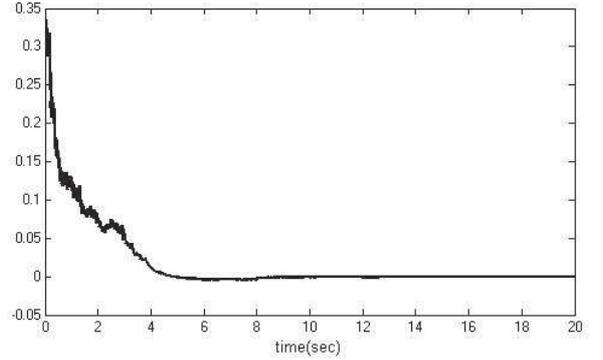}
  \caption{Control input $u$}\label{Fig.2}
\end{figure}
\begin{figure}[H]
\centering
  \includegraphics[width=8cm]{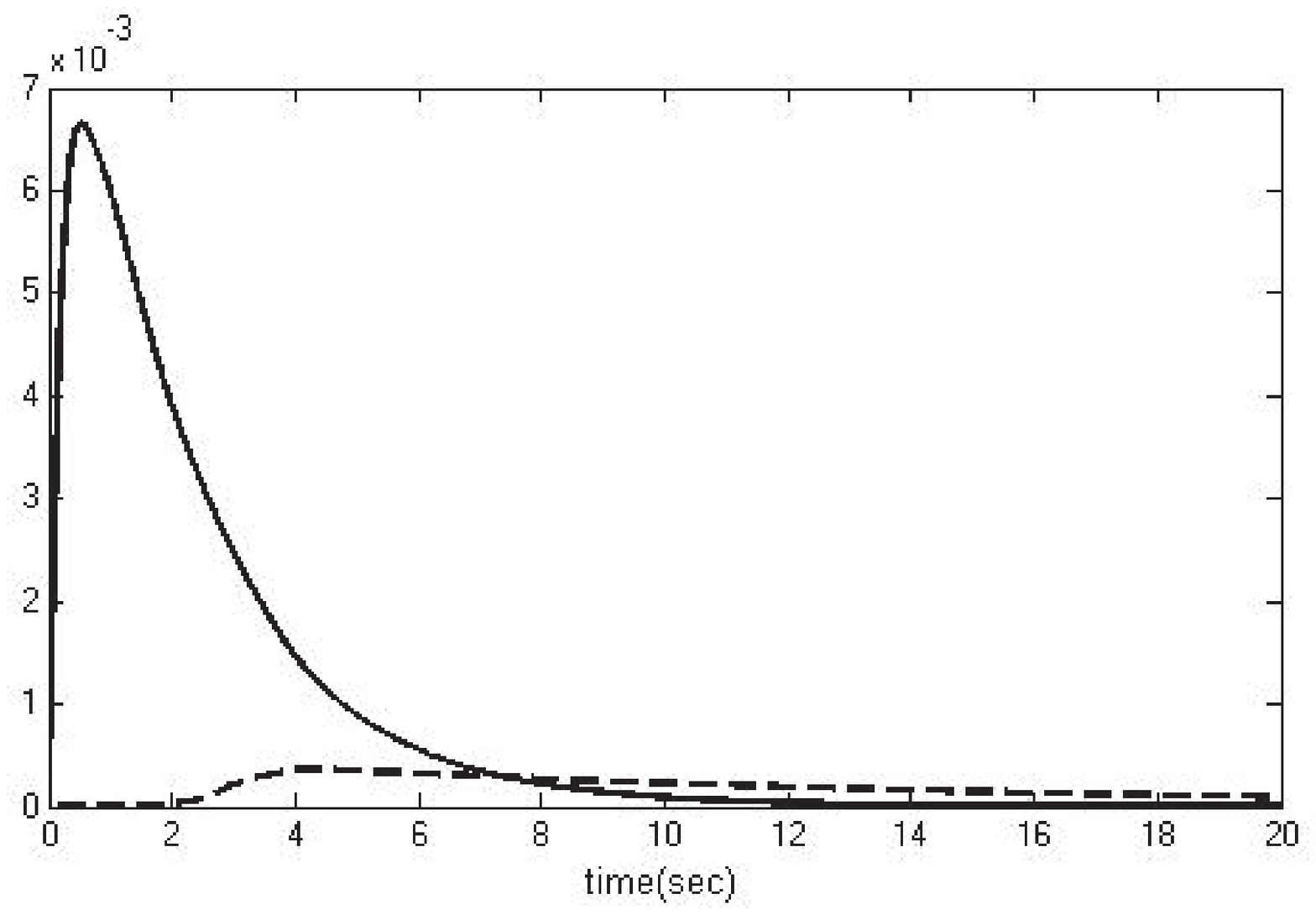}
  \caption{Boundedness of weights $\parallel{\hat{W}_1}\parallel$:``solid line."$\parallel{\hat{W}_2}\parallel$:``dash line".}\label{Fig.3}
\end{figure}
\begin{figure}[H]
\centering
  \includegraphics[width=8cm]{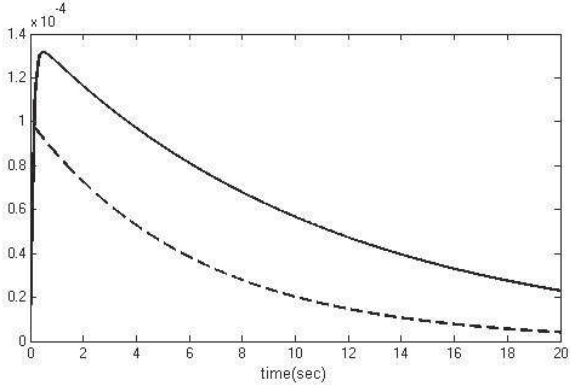}
  \caption{Boundedness of approximation error $\hat{\varepsilon}_1$:``solid line."$\hat{\varepsilon}_2$:``dash line".}\label{Fig.4}
\end{figure}
\begin{figure}[H]
\centering
  \includegraphics[width=8cm]{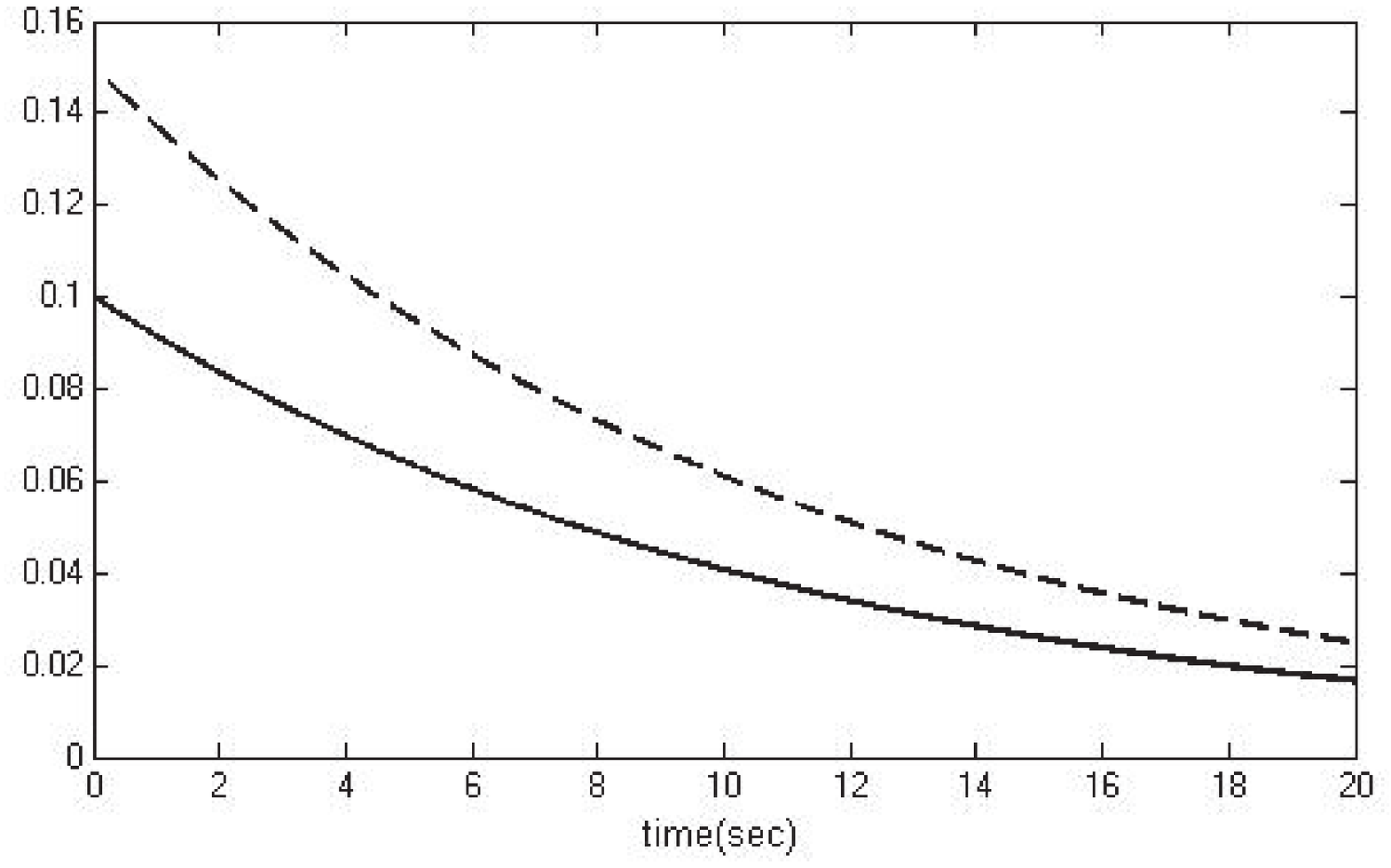}
  \caption{Boundedness of parameters $\parallel\hat{p}\parallel_1$:``solid line."$\parallel\hat{p}\parallel_2$:``dash line".}\label{Fig.5}
\end{figure}
\begin{figure}[H]
\centering
  \includegraphics[width=8cm]{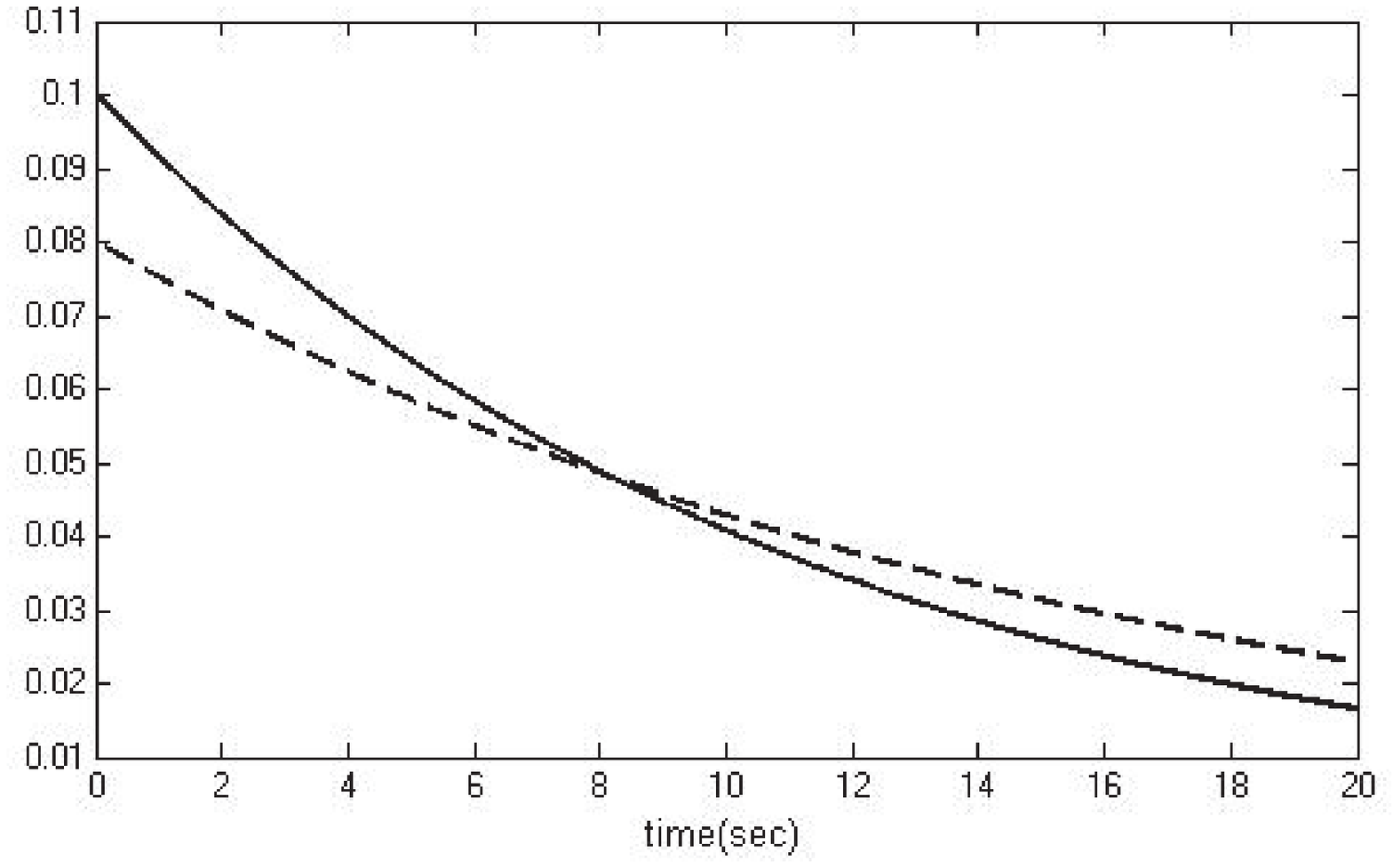}
  \caption{Boundedness of parameters $\parallel\hat{\vartheta}_1\parallel$:``solid line."$\parallel\hat{\vartheta}_2\parallel$:``dash line".}\label{Fig.6}
\end{figure}

The simulation results are shown in Fig.\ref{Fig.1}--Fig.\ref{Fig.6}, from which we can see that the controller renders the resulting closed-loop system asymptotically stable and the limits of estimated parameters exist and are finite. Fig.\ref{Fig.1} shows that $x_1$ and $x_2$ converge to zero rapidly. Fig.\ref{Fig.2} displays the control input signal $u$,  Fig.\ref{Fig.3} shows that boundedness of weights $\|\hat{W}_1\|$ and $\|\hat{W}_2\|$, Fig.\ref{Fig.4} displays the boundedness of approximation error $\hat{\varepsilon}_1$ and $\hat{\varepsilon}_2$, and Fig.\ref{Fig.5} and Fig.\ref{Fig.6} show the response curve of the adaptive parameter $\|\hat{p}_1\|$, $\|\hat{p}_2\|$, $\|\hat{\vartheta}_1\|$ and $\|\hat{\vartheta}_2\|$.

\section{Conclusion}
In this paper, the adaptive neural control method of nonlinear systems is extended to a class of stochastic nonlinear systems with the unknown disturbances, unknown parameters and unknown functions. Unknown nonlinearities can be approximated by RBFNN. All neural network weights are tuned online with no prior training needed. Adaptive bounding design technique is used to deal with the unknown parameters and unknown functions. we have designed a universal type adaptive feedback controller, which keep boundedness of all the states to guarantee global asymptotically stable in probability. Simulation has been conducted to show the performance of the proposed approach.

\def\ack{\section*{Acknowledgements}%
  \addtocontents{toc}{\protect\vspace{6pt}}%
  \addcontentsline{toc}{section}{Acknowledgements}%
}
\ack
This work is partially supported by the National Natural Science Foundation of China (61503133, 51374107, 51577057, 61304162), by Major State Basic Research Development Program (973) sub-project (61325309), and by Postdoctoral Science Foundation of China under Grant (2016M592449), and by Natural Science Foundation of Hunan Province (2016JJ6043, 14JJ3110), and by Research Foundation of Education Bureau of Hunan Province (15C0548).

\end{document}